\documentstyle[12pt,titlepage]{article}

\def\ref{\par\noindent\hangindent=1cm}
\def\mincir{\raise -2.truept\hbox{\rlap{\hbox{$\sim$}}\raise5.truept
\hbox{$<$}\ }}
\def\magcir{\raise -2.truept\hbox{\rlap{\hbox{$\sim$}}\raise5.truept
\hbox{$>$}\ }}
\def\minmag{\raise-2.truept\hbox{\rlap{\hbox{$<$}}\raise 6.truept\hbox
{$>$}\ }}
\def\gr{\kern 2pt\hbox{}^\circ{\kern -2pt K}} 
\begin{document}
\title{}
\author{}
\date{}
\vspace{4cm}
\Huge
\begin{center}
{\bf A Unique Mass Function \protect\\
from Galaxies to Clusters ?}
\end{center}
\vspace{2cm}
\normalsize
\begin{center}
{\em G. Giuricin$^{1,2}$, F. Mardirossian$^{1,2}$,
M. Mezzetti$^{1,2}$ \protect\\
M. Persic$^{2,3}$, \& P. Salucci$^{2}$}
\end{center}
\vspace{0.5cm}
(1) Dipartimento di Astronomia,
Universit\`{a} degli Studi di Trieste, Trieste, Italy \\
(2) SISSA, Strada Costiera 11, I--34014 Trieste, Italy \\
(3) Osservatorio Astronomico, via G.B. Tiepolo 11, I--34131 Trieste, Italy \\
\pagebreak
\renewcommand{\thesection}{\arabic{section}}
\renewcommand{\thesubsection}{\thesection.\arabic{subsection}}
\baselineskip 12pt
\section*{Abstract}
We present an observational mass function ranging from galaxies to massive
galaxy clusters, derived from direct dynamical mass estimates. Our mass
function shows, in the low-mass range of galaxies and groups, a behaviour in
agreement with that of standard CDM ($n=-2$), while in the high-mass range
(clusters) our mass function is shallower (and thus contains more power) than
standard CDM; it also results shallower than the recent mass function
by Bahcall \& Cen (1993).

\vspace{0.5cm}
{\em Subject headings:} \ galaxies: clustering

\section{Introduction}
The observational mass functions of bound systems can provide tight
constraints to theories which describe the Universe density field at different
scales, and, in general, can produce important diagnostics for cosmological
models (see, e.g., Edge et al. 1990; Henry \& Arnaud 1992; Lilje 1992).

Galaxies, groups and clusters of galaxies trace the observed large scale
structure of the Universe, and contribute to map the distribution of the total
(luminous and dark) mass, $M$. In this paper we consider the mass function (MF,
hereafter), $n(M)$, of galaxies, groups and clusters of galaxies; $n(M)$ is the
number of systems per unit volume and per unit mass. These MFs are obtained
straight from the dynamics of galaxies, groups and clusters (see, e.g., Ashman,
Persic, \& Salucci 1993, Pisani et al. 1992, Biviano et al. 1993), rather than
from indirect methods (see, e.g., Bahcall \& Cen 1992; Bahcall \& Cen 1993). In
this way it is possible, in the estimate of the MFs, to avoid using cluster
mass-to-luminosity ratios, which usually translate luminosity functions (LF,
hereafter) into MF, or to avoid being forced to chose some X-ray-sources models
coupled with X-ray-temperature functions (see, e.g., Henry \& Arnaud 1991). In
fact, often indirect techniques are more model-dependent than the direct
analysis is, and may induce biases which are not easily detectable.

In the present paper we use $H_0 = 100\, h$ km s$^{-1}$ Mpc$^{-1}$, and the
masses are expressed in solar units.

\section{The Mass Functions}

We assemble the MF of galaxies, of groups, and of clusters of galaxies, which
have been obtained from direct determinations. The MF, so obtained, is defined
over three available observational windows, corresponding to the mass intervals
of galaxies, groups, and of massive clusters, and it ranges between $\sim 2
\times 10^{11} h^{-1} M_{\odot}$ and $\sim 5 \times 10^{15} h^{-1} M_{\odot}$.
The galaxy MF
is derived from Ashman, Persic, \& Salucci (1993); the group MF from Pisani et
al (1992); and the cluster MF from Biviano et al. (1993).

\subsection {Galaxy MF}
Ashman, Persic, \& Salucci (1993) (see also Persic \& Salucci 1988, 1990) have
shown that, in spiral galaxies, $M \propto L^p$ with $p = 0.5 \pm 0.1$, where
$M$ and $L$ are the halo mass and the luminosity, respectively. They have shown
that, as a consequence of this, at low masses $n(M) \propto M^{-\alpha_{gal}}$,
with $1.6 \mincir \alpha_{gal} \mincir 2$.

In the present paper we assume that the halo MF does not depend on the
morphological type of the embedded optical galaxy, and that the halo mass well
describes the total mass of the galaxy.

In order to normalize the galaxy MF, we use the result (from the dynamics of
galaxy pairs, i.e. independent of optical morphology) that the halo of an
$L_B^*$ galaxy has a mass of $M \sim 1.5 \times 10^{12} h^{-1} M_\odot$
(Charlton \& Salpeter 1991). Using the mixed-morphology $B$-band luminosity
function of Efstathiou, Ellis, \& Peterson (1988) (for $0.03\,L_B^* \mincir L
\mincir 0.5 \,L_B^*$, i.e. in the luminosity range where the luminosity
function is power law), our galaxy MF is the power law
$$
n(M) ~ \simeq ~ 7^{+3.5}_{-2.5} \times 10^7 h^{2.2}~ M^{-1.8 \pm 0.2}
{}~~~~~ {\rm Mpc}^{-3} ~ {\rm M_\odot}^{-1}
\eqno(1)
$$
in the mass range between $2.5 \times 10^{11} h^{-1} M_{\odot}$ and $1.1 \times
10^{12} h^{-1} M_{\odot}$ (corresponding to the above luminosity range).

\subsection{Group MF}
Pisani et al. (1992) have studied the problem of evaluating the distribution of
the masses of galaxy groups by considering some sets of groups mainly belonging
to the Local Supercluster. Out of these sets of groups we have chosen the
groups identified by Tully (1987) in the Nearby Galaxy Catalogue (Tully 1988),
because this catalogue represents a good sample of the nearby Universe,
including also faint (but only gas-rich) galaxies. However, some completeness
problems in the catalogue and the presence of the Local Supercluster in the
sampled volume have suggested us to give only upper- and lower-limit estimates
for the group MF. Following Pisani et al. (1992), we have conservatively chosen
the high-mass groups between $1.1 \times 10^{13} h^{-1} M_{\odot}$ and $1
\times 10^{14} h^{-1} M_{\odot}$. In this mass range the group MF can be well
described by the power law:
$$
n(M) ~ = ~ A_{gr} ~ h^{2} ~ M^{-2}\ {\rm Mpc}^{-3} ~ {\rm M_\odot}^{-1},
\eqno(2)
$$
where $A_{gr}$ assumes the values $3.9 \times 10^{10}$ and $1.4 \times 10^{10}$
for the upper and the lower limit MF, respectively.

In order to estimate $A_{gr}$, we have taken into account the obscuration
produced by our Galaxy (within $|b| \leq 15^o$) and the fraction ($\sim 40 \%$)
of groups comprised in the above-mentioned mass range. The value of $A_{gr}$
corresponding to the MF upper limit has been deduced from the number of groups
contained in the volume where the Nearby Galaxy Catalogue is complete
(corresponding to systemic velocities smaller than 1500 km s$^{-1}$). We
consider this estimate as an upper limit because the volume considered contains
the richest region of Virgo Supercluster. On the other hand, the incompletion
correction factor given by Tully (1987) allows us to estimate the total number
of groups within 3000 km s$^{-1}$ (including also faraway binary galaxies)
which, therefore, leads to a value of $A_{gr}$ corresponding to the MF lower
limit. In fact, a volume corresponding to a radius of 3000 km s$^{-1}$ samples
the Universe better than a volume of radius $1500 km s^{-1}$ and, moreover,
faint and gas-poor galaxies belonging to faraway groups are probably lost thus
reducing the number of identified systems.

\subsection{Cluster MF}
Biviano et al. (1993) have studied the distribution function of the masses of
75 clusters, each having at least 20 galaxy members with measured redshifts
within $1.5\,h^{-1}$ Mpc, and with mean redshift $z \leq 0.15$ (in order to
reasonably reduce evolutionary effects). Possible subclustering problems have
suggested to consider only the masses evaluated, via the Virial Theorem, within
an aperture of half the Abell radius ($0.75 \,h^{-1}$ Mpc) in 69 clusters.
Subsequently, problems of incompleteness in the medium/low-mass regime have
suggested to conservatively define only the MF of massive clusters. This can be
represented by the power law:
$$
n(M) ~ \simeq ~ 1.0 \times 10^{14} h^{1.7} ~M^{-2.3}\
{\rm Mpc}^{-3} ~ {\rm M_\odot}^{-1}
\eqno(3)
$$
in the mass range $4 \times 10^{14} h^{-1} M_{\odot} \mincir M \mincir 1.6
\times 10^{15} h^{-1} M_{\odot}$. The numerical coefficient of Eq.3 (see
Biviano et al. 1993) corresponds to Peacock \& West (1992) estimate of the
number density of clusters (this value is $\sim 1.5$ times smaller than those
reported by Scaramella et al. 1992 and Zucca et al. 1993). The associated
uncertainty band has been estimated by means of the error propagation method.

Girardi et al. (1993) have shown that most of the clusters considered are quite
well described by King density profiles. The median and mean cluster core radii
are $\sim 0.15 \,h^{-1}$ Mpc and $\sim 0.17 \,h^{-1}$ Mpc, respectively. We
adopted a multiplicative coefficient $3$ (in the masses) to translate Biviano
et al.'s (1993) original MF (obtained with masses evaluated within half the
Abell radius, i.e. typically $\sim 4-5$ core radii) into an asymptotic MF. So,
Eq.(3) (with Peacock \& West's density) can be considered as a lower limit to
the cluster MF; vice-versa, the MF with masses $3$ times as large can be
considered as an upper limit.

\section{A Unique Mass Function ?}

Eqs.(1), (2), and (3) describe the MF of galaxies, of massive groups of
galaxies, and of massive clusters of galaxies, respectively. It is possible to
use these functions and their uncertainty bands to constrain a unique MF of
bound systems which could give the mass distribution from galaxies up to
massive clusters. Fig.1 shows these MFs together with the respective
uncertainty bands (for groups, only the upper and lower limits are drawn).

The presence of MF-uncertainty-bands suggests to consider for the putative
unique MF both an upper limit and a lower limit, if an analytic function is
used to fit the logarithmic data. On the other hand, if a linear fit is
preferred, it may be useful to obtain the steepest and the shallowest lines
allowed within the uncertainty bands, and to consider these two lines as limits
for the MF.

We use a Press-Schechter function:
$$
n(M)~=~ A ~h^{4} ~ (M/M^{*})^{\alpha - 2} exp[-(M/M^{*})^{2 \alpha}]\
{\rm Mpc}^{-3} ~ {\rm M_\odot}^{-1},
\eqno(4)
$$
as the analytic function to fit our logarithmic data. The upper and the lower
limits are defined by the parameter sets ($A= 1.0 \times 10^{-19}, M^{*}= 1.0
\times 10^{15} h^{-1} M_{\odot}, \alpha=0.2$) and ($A= 1.3 \times 10^{17},
M^{*}= 5.0 \times 10^{13} h^{-1} M_{\odot}, \alpha=0.2$), respectively.

On the other hand, if we try a linear fit to our logarithmic data, the two
lines giving the upper limit and the lower limit virtually coincide, with
slopes ranging between $-1.9$ and $-2.1$. Therefore we give a single power
law:
$$
n(M)~=~ 1.5 \times 10^{10} ~h^{2} ~M^{-2}\
{\rm Mpc}^{-3} ~ {\rm M_\odot}^{-1}.
\eqno(5)
$$
In Fig.2 we plot the upper and the lower limit to the unique MF according to
Eq.(4), as well as its linear representation given in Eq.(5).

Our observational MF shows, in the low-mass range of galaxies and groups, a
behaviour in agreement with that of standard CDM ($n=-2$) while at high masses
(clusters) our MF is shallower (and thus contains more power) than standard
CDM. It also results shallower than the cluster MF recently derived by Bahcall
\& Cen (1993).

In the hypothesis that the fraction $f$ of galaxies contained in systems
(groups and clusters) is equal to the ratio between the total mass-densities
due to systems and due to galaxies, we can obtain the range of values within
which the mean mass density in bound objects, $\Omega_{bound}$, is expected to
be. The values of $\Omega_{bound}$, estimated for the range of masses between
$1 \times 10^{11} h^{-1} M_{\odot}$ and $1 \times 10^{16} h^{-1} M_{\odot}$,
are $0.37 (1+f)^{-1}$, $1.34 (1+f)^{-1}$, $0.65 (1+f)^{-1}$, for the lower and
the upper MF limits, and for the power-law MF fit, respectively. This implies
$\Omega_{bound} \sim 0.5$ for $f=0.7$, which is thought to be the value
tipical for the nearby Universe (see, e.g., Tully 1987).
\vspace{0.5cm}

This work has been partially supported by the {\em Ministero per l'Universit\`a
e per la Ricerca Scientifica e Tecnologica}, and by the {\em Consiglio
Nazionale delle Ricerche (CNR-GNA)}.

\section*{References}

\ref Ashman, K.M. 1992, PASP, 104, 1109.

\ref Ashman, K.M., Salucci, P., \& Persic, M., 1993, MNRAS, 260, 610.

\ref Bahcall, N.A., \& Cen, R. 1992, ApJ, 398, L81.

\ref Bahcall, N.A., \& Cen, R. 1993, ApJ, 407, L49.

\ref Biviano, A., Girardi, M., Giuricin, G., Mardirossian, F., \& Mezzetti, M.
1993, ApJ L, in press.

\ref Borgani, S., Bonometto, S., Persic, M., \& Salucci, P. 1991, ApJ, 374, 20.

\ref Charlton, J.C., \& Salpeter, E.E. 1991, ApJ, 375, 517.

\ref Edge, A.C., Stewart, G.C., Fabian, A.C., \& Arnaud, K.A. 1990, MNRAS,
245, 559.

\ref Efstatiou, G., Ellis, R.S., \& Peterson, B.A. 1988,
MNRAS, 232, 431.

\ref Girardi, M., Biviano, A., Giuricin, G., Mardirossian, F., \&
Mezzetti, M., 1993, in preparation.

\ref Henry, J.P., \& Arnaud, K.A. 1991, ApJ, 372, 410.

\ref Lilje, P.B. 1992, ApJ, 386, L33.

\ref Peacock, J.A., \& West, M.J. 1992, MNRAS, 259, 494.

\ref Persic, M., \& Salucci, P. 1988, MNRAS, 234, 131.

\ref Persic, M., \& Salucci, P. 1990, MNRAS, 245, 577.

\ref Pisani, A., Giuricin, G., Mardirossian, F., \& Mezzetti, M.
1992, ApJ, 389, 68.

\ref Scaramella, R., Zamorani, G., Vettolani, G., \& Chincarini,
G. 1991, AJ, 101, 342.

\ref Tully, B. 1987, ApJ, 321, 280.

\ref Tully, B. 1988, {\em Nearby Galaxy Catalog}, Cambridge
University Press, Cambridge, U.K.

\ref Zucca, E., Zamorani, G., Scaramella, R., \& Vettolani, G.
1993, ApJ, in press.

\section*{Captions to Figures}
{\bf Fig.1:}
The MFs together with the respective uncertainty bands.

\noindent {\bf Fig.2:}
The upper and the lower limits for the
unique MF, and the power law description of the unique MF.

\end{document}